# Mitrion-C Application Development on SGI Altix 350/RC100

Volodymyr V. Kindratenko, Robert J. Brunner, and Adam D. Myers
*National Center for Supercomputing Applications, University of Illinois at Urbana-Champaign*
*kindr@ncsa.uiuc.edu, rb@astro.uiuc.edu, adm@astro.uiuc.edu*

## Abstract

*This paper provides an evaluation of SGI® RASC™ RC100 technology from a computational science software developer's perspective. A brute force implementation of a two-point angular correlation function is used as a test case application. The computational kernel of this test case algorithm is ported to the Mitrion-C programming language and compiled, targeting the RC100 hardware. We explore several code optimization techniques and report performance results for different designs. We conclude the paper with an analysis of this system based on our observations while implementing the test case. Overall, the hardware platform and software development tools were found to be satisfactory for accelerating computationally intensive applications, however, several system improvements are desirable.*

## 1. Introduction

High-performance reconfigurable computing (HPRC) based on the combination of conventional microprocessors and field-programmable gate arrays (FPGAs) is growing beyond its original niche. This technological approach combines the advantages of the coarse-grain, process-level parallel processing provided by conventional multiprocessor systems with the fine-grain, instruction-level parallel processing available with FPGAs [1]. HPRC in its present stage is a rapidly evolving technology with interesting potential, but it has yet to overcome numerous challenges in order to be accepted as a mainstream computing paradigm. Programmability, code compatibility, and portability are among the challenges the technology currently faces before its acceptance by the computational community.

Traditional high-performance computing (HPC) vendors such as SGI® and Cray Inc. have introduced several commercial HPRC products. In addition, a number of newcomers in the HPC arena including SRC Computers, Inc. and Nallatech Ltd. have emerged with their own solutions. Some vendors provide both hardware and software solutions, where others focus only on the hardware or software side of the technology.

SGI Reconfigurable Application-Specific Computing (RASC™) architecture [2, 3] is a prime example of a heterogeneous HPRC system in which traditional microprocessors and FPGAs can be used together to accelerate computationally demanding applications. SGI provides FPGA-based hardware and a supporting software stack that enables an application to interact with the reconfigurable hardware. Tools for implementing an actual algorithm design for the FPGA hardware, however, are left to third party solution providers.

Traditionally, FPGA designs have been built using hardware description languages, such as VHDL or Verilog. In recent years, however, several tools have been developed that compile a code written in a high-level language directly into a hardware circuitry description, examples include Handel-C, Impulse C, MAP C, and Mitrionics™ Mitrion-C SDK [4]—the tool used in this study. The source code written in Mitrion-C is compiled by the Mitrion-C compiler into a code for the Mitrion processor—a proprietary virtual software processor—followed by an automatic adaptation and implementation of the processor in the FPGA, which targets a specific hardware platform, such as the SGI RASC. The traditional hardware synthesis, map, place, and route tools, such as those provided by Xilinx Inc., can be used to generate the FPGA configuration file. This approach completely eliminates the need for a low-level hardware design, and makes it relatively straightforward for an application software developer to implement algorithmic cores on the RC100 platform.

In this report, we provide a technical analysis of SGI's third generation RASC system, the RC100. The evaluation of this system is based on the implementation of the kernel of a computationally intensive algorithm used to study the distribution of matter in the universe [5]. The algorithm is computationally and data intensive, and, therefore, it provides a good test case that requires double-precision floating-point arithmetic with computational complexity of $O(N^2)$. We use the Mitrion-C programming language to develop the FPGA-accelerated side of the algorithm, while the host CPU code is implemented in standard ANSI C using the RASC library [3]. We compare the performance of two FPGAs from the RC100 blade to the performance of two 1.4 GHz Intel Itanium 2 chips on the SGI Altix 350 system. Table 1 lists the versions of the different software components we used in the course of this work.

| Tool | Version |
|---|---|
| SGI RASC | RASC 2.0 SGI ProPack 4 Service Pack 3 |
| Mitrion SDK | 1.2.3 build 224 |
| Xilinx ISE | 8.1.03i |
| gcc | 3.2.3 |

**Table 1.** The specific versions of the software tools used in this analysis.

This paper is organized as follows. We describe the SGI RC100 hardware architecture and RASC software stack in Section 2. The RASC software development flow using Mitrion SDK is reviewed in Section 3. We present the test case algorithm in Section 4. In Section 5 we detail our evaluations, including: details of several algorithm implementations on RC100 platform, various optimization techniques, and performance results and comparison with a related CPU-based implementation. Finally, we discuss our evaluation of the system based on our experience in porting our test case in Section 6.

## 2. The SGI Altix 350 with RC100 blade

In our analysis, we use a standalone, single-module SGI Altix 350 system [6] with a single dual-blade chassis containing one RC100 blade [3]. The SGI Altix 350 is a dual-1.4 GHz Intel Itanium 2 system with 4 GBs of physical memory. An RC100 blade is attached to the host system via a NUMAlink 4 interconnect (see Fig. 1).

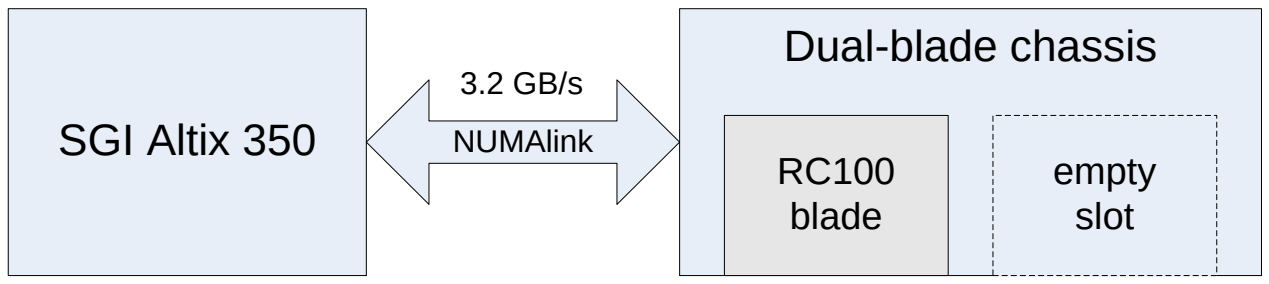

**Figure 1.** SGI Altix 350 with dual-blade chassis containing one RC100 blade.

The RC100 is SGI's third-generation Reconfigurable Application-Specific Computing hardware module. It contains two computational FPGAs, two peer-to-peer I/O (TIO) ASICs and a special-purpose FPGA for loading bitstreams onto the computational FPGAs (see Fig. 2). The two user FPGAs are connected to the corresponding TIO ASICs via the Scaleable System Ports (SSPs). In addition, 10 QDR SRAM memory modules, each up to 8 MB, can be installed, in configuration up to five banks per FPGA chip. The SRAMs are configured as two banks to match the NUMAlink 4 channel bandwidth (3.2 Gbyte/sec) to the memories (2x1.6 Gbyte/sec).

The two user FPGAs are Xilinx Virtex 4 LX200 (XC4VLX200-FF1513-10) chips. Each chip contains 200,448 logic cells, 336 Block RAM/FIFOs with 6,048 kbits of total Block RAM, 96 DSP48 slices, and 960 user I/O pins. The maximum clock frequency of the chips, as implemented in the RC100, is 200 MHz. A portion of each chip is allocated to the RASC Core Services logic with the rest of the logic allocated to the user algorithm block (see Fig. 3).

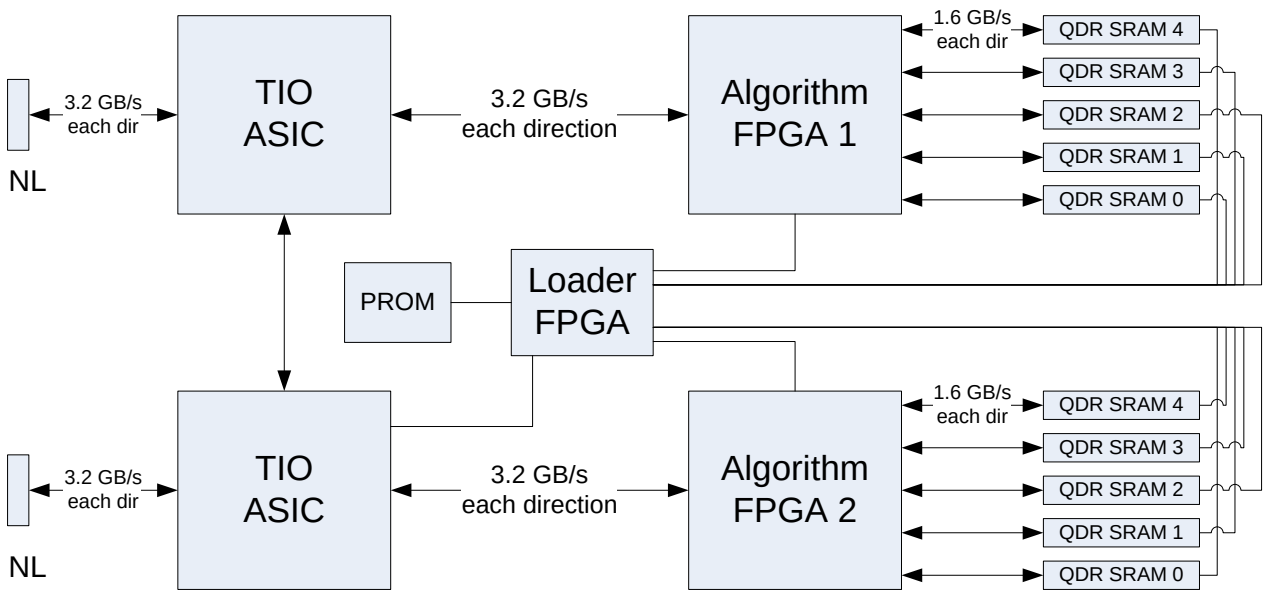

**Figure 2.** A diagram showing the RC100 blade hardware.

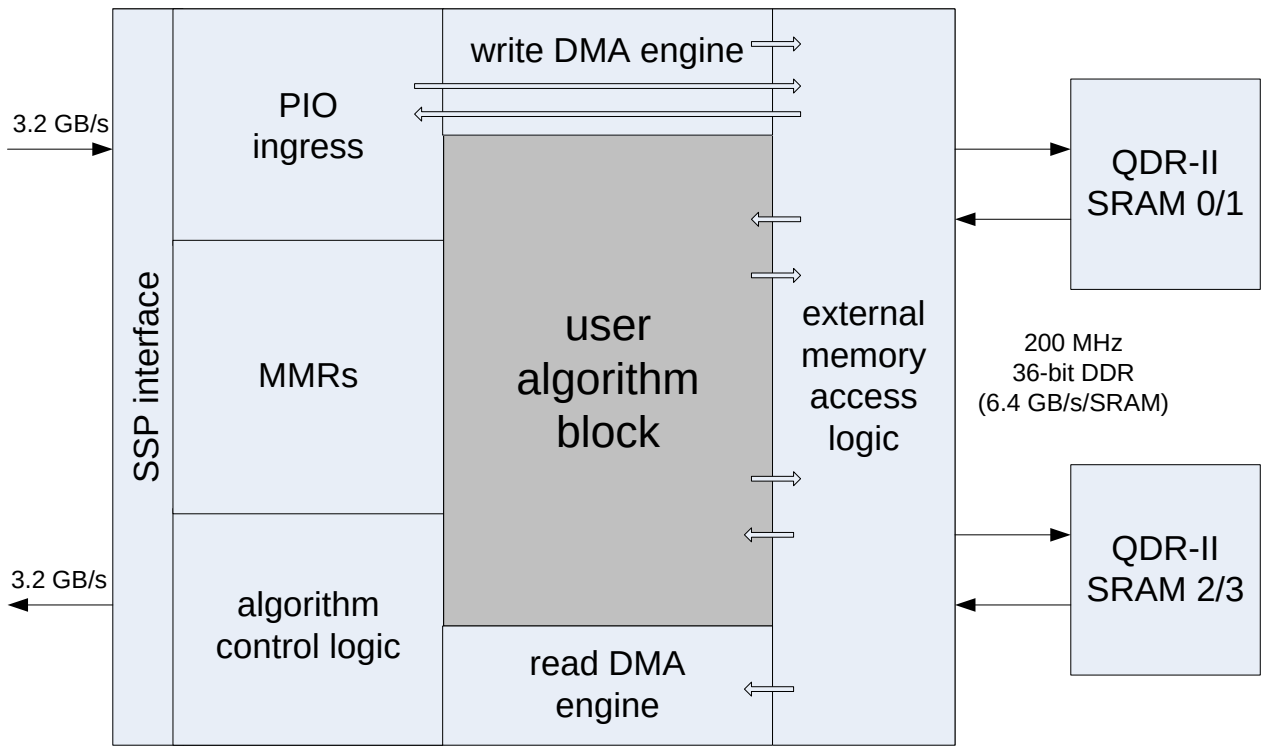

**Figure 3.** A diagram showing the computational FPGA with the User Algorithm Block and various components of the RASC Core Services.

RASC Core Services implement an interface between the TIO chip and SRAMs attached to the FPGA. They also provide memory-mapped register (MMR) inputs and algorithm control logic. The user algorithm block has direct access to MMR inputs and access to SRAM via external memory logic implemented as part of the RASC Core Services. The user algorithm block has access to two dual-port memory resources; each is 128-bit wide and up to 1 million words deep (16 MB per port per bank).

We used the RASC 2.0 SGI ProPack 4 Service Pack 3 software, which is detailed in Fig. 4. The RASC kernel driver and FPGA bitstream download driver operate at the OS level. The RASC Abstraction Layer provides an API for both the kernel device driver and the RASC hardware. This software layer implements data movement from/to devices and spreads the workload across multiple devices. The Abstraction Layer is implemented as two layers: the Co-Processor (COP) level and the algorithm level which is built on top of the COP layer. The COP layer provides an interface to work with individual devices, whereas the algorithm layer treats a collection of devices as a single logical device. The use of these two layers is mutually exclusive. User applications make calls to the RASC Abstraction Layer to gain access to the hardware. The Device Manager is a user-space utility for bitstream management.

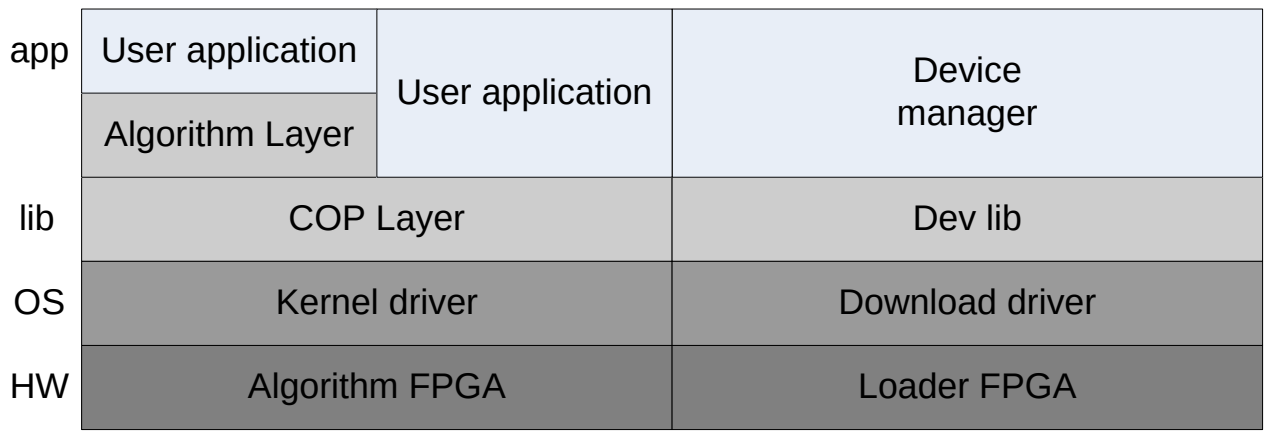

**Figure 4.** The RASC software stack.

## 3. The RASC software development flow

The Mitrion SDK [4] provides the framework in which we implemented our algorithm on the RC100 platform. The bitstream development path using Mitrion-C as a high-level language is shown in Fig. 5. Mitrion-C source can be verified for correctness using a functional simulator/debugger provided with the SDK. The Mitrion-C compiler will generate VHDL code from the Mitrion-C source and setup the instance hierarchy of the RASC FPGA design [3] that includes the user algorithm implementation, the RASC Core Services, and configuration files necessary to implement the design. The design is then synthesized using the Xilinx suite of synthesis and implementation tools (in our testing we used version 8.1.03i). In addition to the bitstream generated by the Xilinx ISE, two configuration files are created: one describes the algorithm's data layout and streaming capabilities to the RASC Abstraction Layer (bitsream configuration file) and the other describes various parameters used by the RASC Core Services. These files, together with the bitstream file, are required by the device manager to communicate with the algorithm that is implemented on the FPGA. Various design verification and debugging capabilities exist in the SDK; however, they are beyond the scope of our work.

We used the Mitrion SDK version 1.2.3, build 224 in our testing, which consists of the Mitrion-C language compiler, integrated development environment, data-dependency graph visualization and simulation tool, Mitrion Host Abstraction Layer (MITHAL) library, and the target platform-specific processor configurator. Mitrion-C source code is compiled into an intermediate virtual processor machine code that can be used by the simulator/debugger or processed by a processor configurator to produce a VHDL design of the application-specific Mitrion Virtual Processor for the final hardware platform (see Fig. 6). Note that the Mitrion Virtual Processor for RC100 is designed to operate at 100 MHz, half the max clock frequency of the RC100 blade.

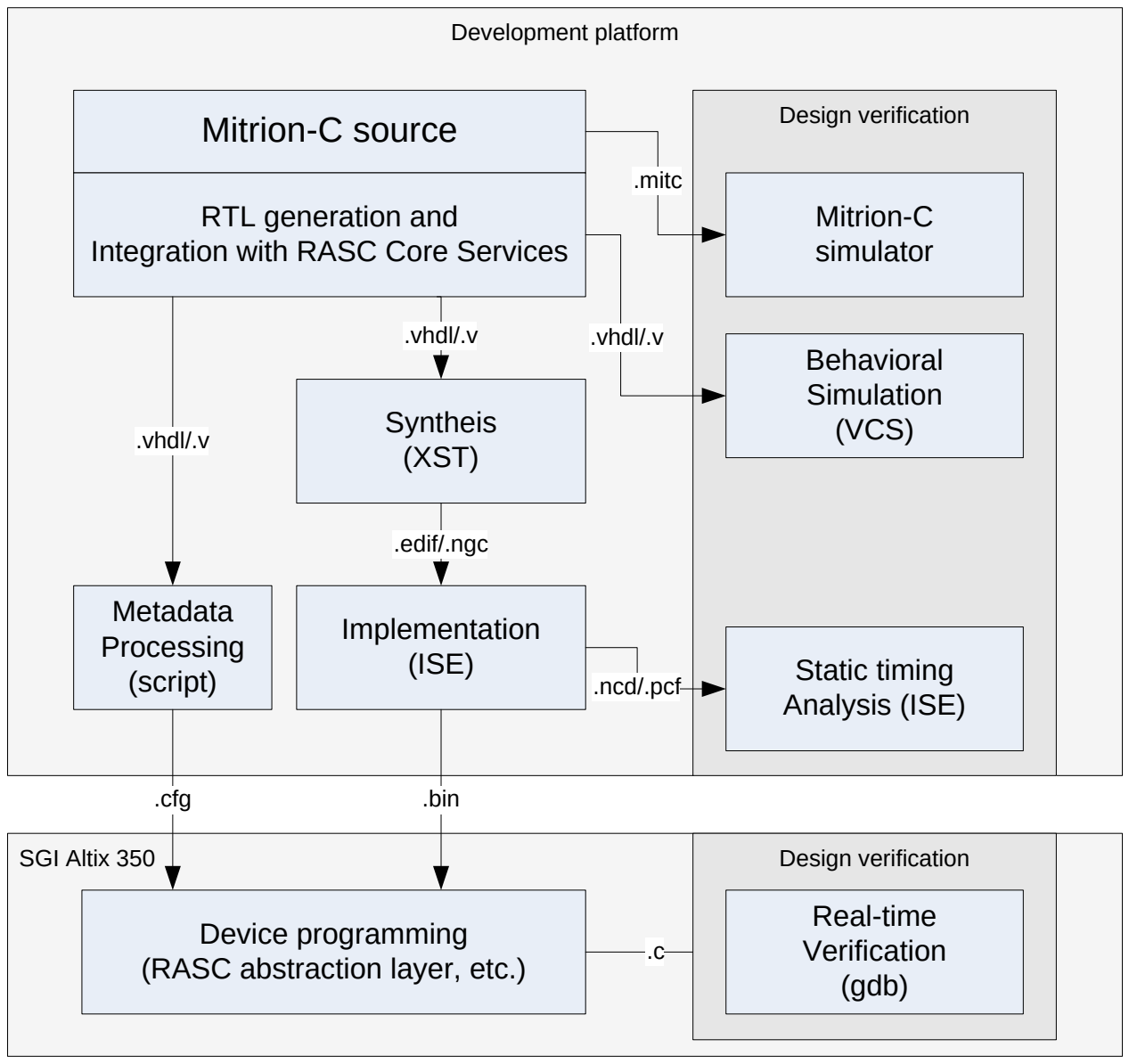

**Figure 5.** The RASC FPGA algorithm design flow.

The Mitrion-C programming language [4] is an intrinsically parallel language. Its data types, such as lists and vectors, and language constructs, such as loops, are designed to support parallel execution driven by the data dependencies. The Mitrion Virtual Processor [4], a proprietary product, is a parallel soft-core processor for FPGAs that executes software written in the Mitrion-C programming language. In essence, it is a dataflow graph in which each node implements a processing element specific to the user application ("cluster on a chip").

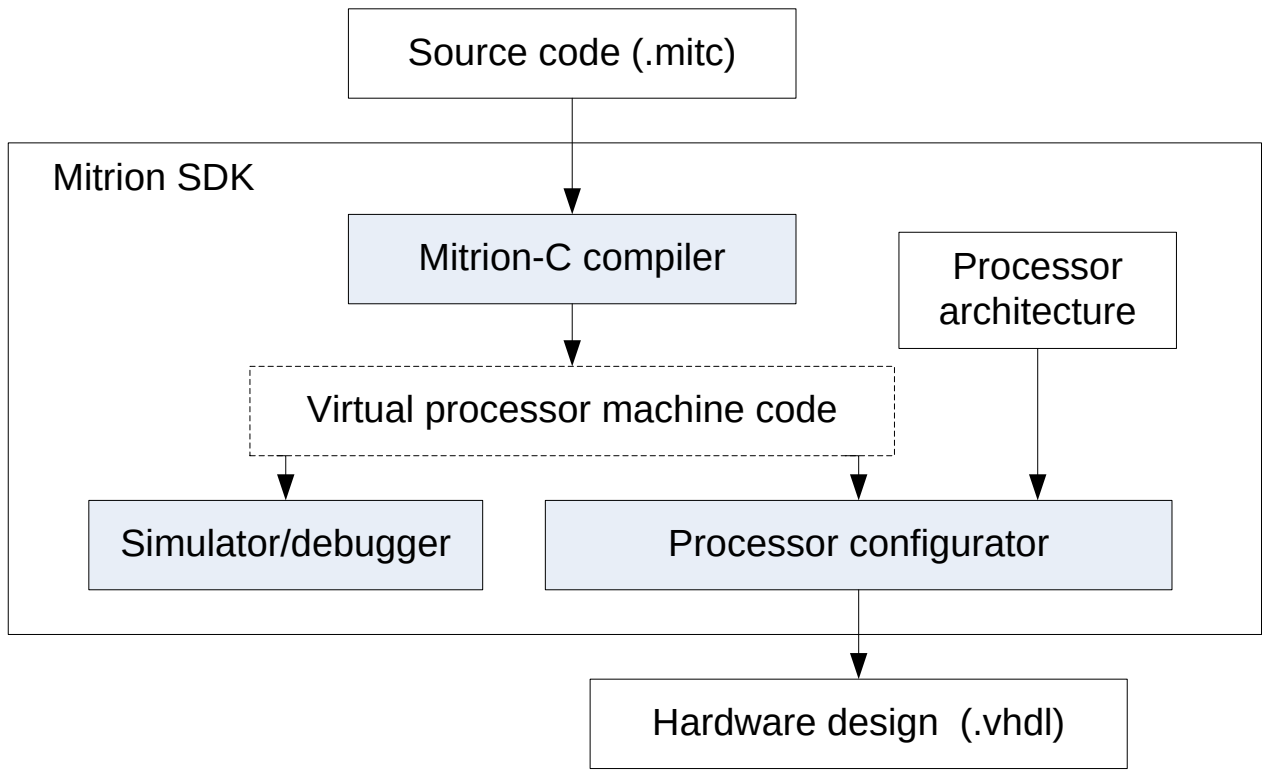

**Figure 6.** The Mitrion-C compilation process.

## 4. The test case

As a test case, we consider the problem of computing the two-point correlation function [7]. The two-point correlation function measures the frequency distribution of separations between coordinate positions in a parameter space as compared to randomly distributed coordinate positions across the same space. In this work, we focus on correlation functions that characterize the clustering of extragalactic objects [e.g., 5] as they are both computationally and data intensive. We focus on the angular separations, $\theta$, between two objects on the celestial sphere, which form the basis for the two-point angular correlation function (TPACF), which we denote $\omega(\theta)$. Qualitatively, a positive value of $\omega(\theta)$ indicates that objects are found *more frequently* at angular separations of $\theta$ than would be expected for a randomly distributed set of coordinate points (a *correlation*).

A detailed description of the underlying mathematical model used to compute TPACF can be found in [7] and [8]; a brief summary is provided below. The two-point angular correlation function is computed as

$$\omega(\theta) = \frac{n_R \cdot DD(\theta) - 2\sum_{i=0}^{n_R-1} DR_i(\theta)}{\sum_{i=0}^{n_R-1} R_i R_i(\theta)} + 1 \qquad (1)$$

where $n_R$ is the number of sets of random points, *DD*, *DR*, and *RR* are angular separation distributions between data points, data and random points, and random points, respectively. The random data sets are equivalent in size to the data set under analysis.

Observationally, determining $\omega(\theta)$ requires binning the separation distributions at some angular resolution $\Delta\theta$. The binning schema used by astronomers is typically logarithmic, as clustering patterns can be important in extragalactic astronomy across a wide range of angular scales. Each decade of angle in the logarithmic space is divided equally between *k* bins, meaning that there are *k* equally-logarithmically-spaced bins between, for example, 0.01 and 0.1 arcminutes. The bin edges are then defined by $10^{j/k}$, where $j=-\infty,\ldots,-1,0,1,\ldots+\infty$ Thus, the problem of computing angular separation distributions can be expressed as follows:

- Input: Set of points $x_1$, .., $x_n$ distributed on the surface of a sphere, and a small number *M* of bins: $[\theta_0, \theta_1)$, $[\theta_1, \theta_2)$, .., $[\theta_{M-1}, \theta_M]$.
- Output: For each bin, the number of unique pairs of points $(x_i, x_j)$ for which the angular distance is in the respective bin: $B_l = |\{ij: \theta_{l-1} <= x_i \cdot x_j < \theta_l\}|$.

The computation of the angular distance $\theta$ between a pair points on the sphere requires converting the spherical coordinates to Cartesian coordinates, computing their dot product, and taking the arccosine of the computed dot product. Once the distance is known, it can be mapped into the respective angular bin $B_l$:

$$l = \mathrm{int}\left[k(\log_{10}\theta - \log_{10}\theta_{\min})\right] \qquad (2)$$

where $\theta_{min}$ is the smallest angular separation that can be measured. Note that this binning schema requires the calculation of the *arccosine* and *logarithm* functions, which are computationally expensive. If only a small number of bins are required, a faster approach is to project the bin edges to the pre-arccosine "dot product" space and search in this space to locate the corresponding bin. Since the bin edges are ordered, an efficient binary search algorithm can be used to quickly locate the corresponding bin in just $log_2M$ steps. We therefore adopt this approach to determine the binned counts.

Henceforth, we will refer to DD($\theta$) or RR($\theta$) counts as autocorrelations and DR($\theta$) counts as cross-correlations, and the full angular two-point autocorrelation, as defined by Equation 1, as the TPACF, or simply as $\omega(\theta)$. Note that formally, the calculation of the cross-correlation requires $N_D^2 log_2M$ steps whereas the autocorrelation is computed in $(N_D(N_D-1)/2)log_2M$ steps.

We use a sample of photometrically classified quasars, and random catalogs, first analyzed by [5] to calculate $\omega(\theta)$. We specifically used one hundred random samples ($n_R=100$); the actual dataset and each of the random realizations contain ~100,000 points ($N_D$=97160). In addition, we employed a binning schema with five bins per decade (*k=5*), $\theta_{min}=0.01$ arcminutes and $\theta_{max}=10000$ arcminutes. Thus, angular separations are spread across 6 decades of scale and require 30 bins (*M=30*). Covering this range of scales requires the use of double-precision floating-point arithmetic as single-precision numbers do not provide the precision sufficient to compute $\theta$ values smaller than 1 arcminute.

## 5. The TPACF on the SGI Altix/RC100

### 5.1. Reference C implementation of the TPACF

As a reference, we first consider a C implementation of the algorithm presented in Section 4. The computational core of the algorithm is a subroutine which calculates binned separation distributions for either DD($\theta$) or DR($\theta$) binned counts, depending on the input parameters. Pseudo code of the core of this subroutine can be found in the Appendix Code 1.

Initially, the data points are loaded and converted from spherical to Cartesian coordinates, and the DD($\theta$) autocorrelation is computed. Next, $n_R$ sets of $N_D$ random points are loaded/converted one set at a time. For each random set, the RR($\theta$) autocorrelation and the DR($\theta$) cross-correlation are computed and stored. Finally, we compute $\omega(\theta)$ according to Equation 1. Pseudo code of the overall algorithm can be found in the Appendix Code 2.

The SGI Altix system contains two processors; thus the cross-correlation and autocorrelation subroutines can be executed simultaneously. In the reference C implementation, we use pthreads to implement the parallel execution of these code sections. The data load and conversion subroutine is kept out of the parallel code to enable a more precise performance characterization. We present performance characteristics of this implementation in the second Column Table 6. This code was compiled with the gcc version 3.2.3 compiler that was supplied with the SGI Altix 350 system using the -O3 -ffast-math -funroll-loops -fprefetch-loop-arrays optimizations. The overall execution time of the reference C implementation, including data I/O, was 86,315.7 seconds.

### 5.2. Mitrion-C kernel implementation

The autocorrelation/cross-correlation subroutine was re-written in the Mitrion-C language, targeting the RC100 platform. Structurally, the Mitrion-C implementation of the computational core, which is presented in Appendix Code 3, resembles the reference C implementation. The Mitrion-C data dependency graph produced by the Mitrion-C simulator is shown in Fig. 7. This graph can be used to verify the run-time behavior of the implementation before it is compiled into hardware. Thus, on each iteration of the outer loop, a new point is loaded from the off-chip memory and is used throughout the entire inner loop execution. On each iteration of the inner loop, a new point is loaded from the off-chip memory and is used in the computation of the dot product. Once the dot product is computed, the bin to which it belongs is identified and updated. Actual bin boundaries are hardcoded in this initial implementation; they are saved as a vector which is stored on the chip. This storage mechanism allows the Mitrion-C compiler to fully unroll the bin array search 'for' loop into a 32-stage *deep* pipeline. Once the bin index is found, the corresponding bin value is incremented by one. Initially, bin values are stored as a vector and set to zero. As with the bin boundaries, this storage mechanism allows the Mitrion-C compiler to fully unroll the bin update 'foreach' loop into a *wide* pipeline. Since the bin search and bin update loops can be fully unrolled, the compiler is able to produce a fully pipelined inner loop implementation, thus generating an efficient overall algorithm implementation in which a new result is produced on each clock cycle. After all the calculations are done, the resulting bin values are written back to the off-chip memory. From there, they are copied to the host memory via a RASC library call.

On the RC100 platform, the Mitrion-C processor has access to just two off-chip memory banks; each such bank is 128-bit wide and a few Megabytes deep. This memory is single-ported as far as the memory read access from the user application is concerned. Since the point coordinates are stored as double-precision floating-point numbers, each point requires 3x64 bits of storage space. In order to avoid a pipeline stall while reading each data point, we distribute coordinate points between the two memory banks as shown in Tables 2 and 3. This data storage schema allows simultaneous access to the coordinate values of a single data point within a single clock.

| | | Bank 1 | | Bank 2 | |
|---|---|---|---|---|---|
| | Memory address | 0…63 bits | 64…127 bits | 0…63 bits | 64…127 bits |
| data set 1 | 0 | $x_0$ | $y_0$ | $z_0$ | unused |
| | *i* | $x_i$ | $y_i$ | $z_i$ | unused |
| | N-1 | $x_{N-1}$ | $y_{N-1}$ | $z_{N-1}$ | unused |
| data set 2 | N | $x_0$ | $y_0$ | $z_0$ | unused |
| | N+*j* | $x_j$ | $y_j$ | $z_j$ | unused |
| | 2N-1 | $x_{N-1}$ | $y_{N-1}$ | $z_{N-1}$ | unused |

**Table 2.** Off-chip memory usage for the cross-correlation calculation. Here N denotes the number of points in each dataset (same as $N_D$ in Chapter 4 and NPOINTS in the Appendix Code 3).

| | | Bank 1 | | Bank 2 | |
|---|---|---|---|---|---|
| | Memory address | 0…63 bits | 64…127 bits | 0…63 bits | 64…127 bits |
| data set 1 | 0 | $x_0$ | $y_0$ | $z_0$ | unused |
| | *i* | $x_i$ | $y_i$ | $z_i$ | unused |
| | N-1 | $x_{N-1}$ | $y_{N-1}$ | $z_{N-1}$ | unused |
| data set 1 | N | $x_0$ | $y_0$ | $z_0$ | unused |
| | N+*i* | $x_i$ | $y_i$ | $z_i$ | unused |
| | 2N-1 | $x_{N-1}$ | $y_{N-1}$ | $z_{N-1}$ | unused |

**Table 3.** Off-chip memory usage for the autocorrelation calculation. The same dataset is repeated twice in the off-chip memory; this means we can use the same subroutine for both the autocorrelation and cross-correlation calculations.

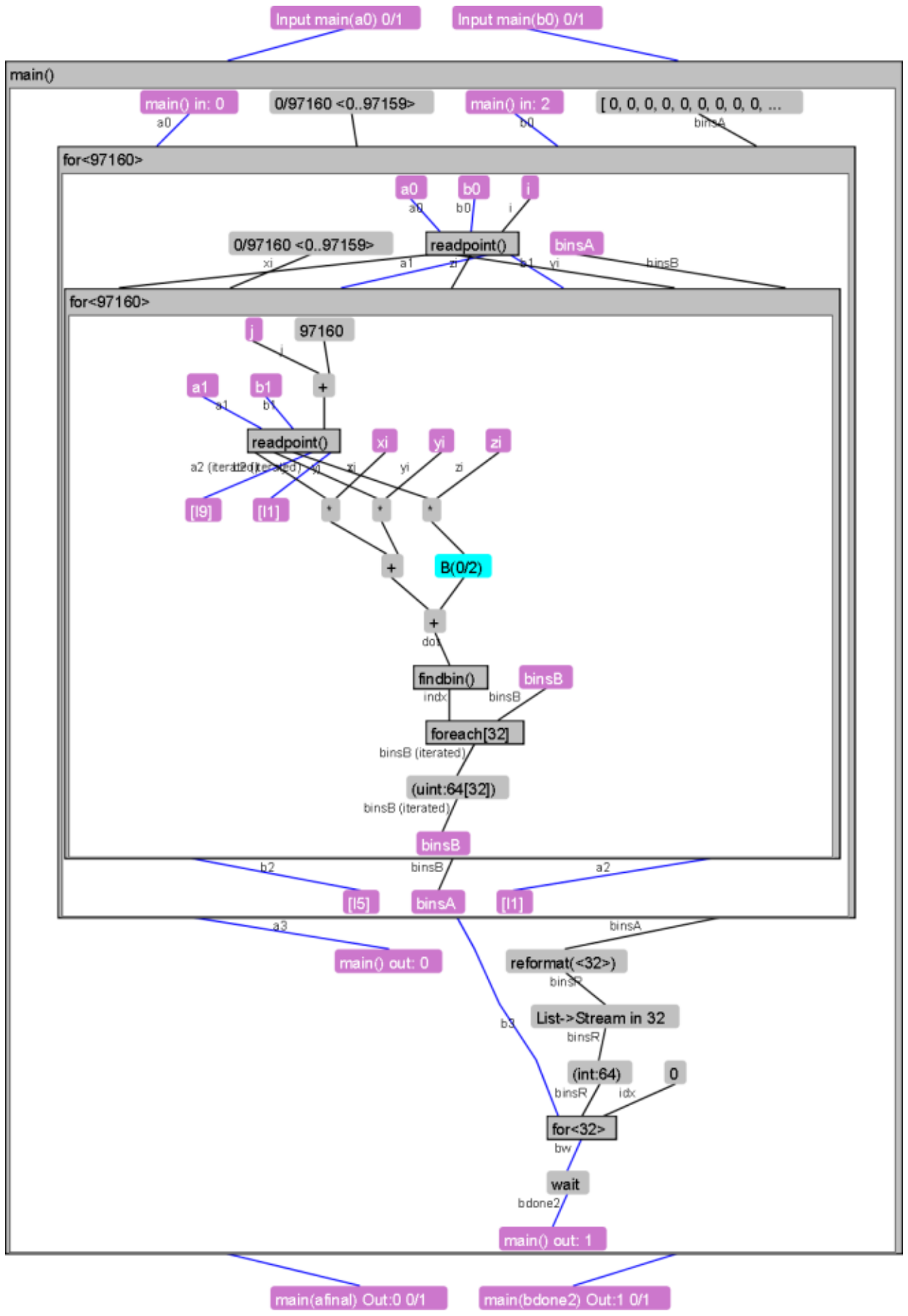

**Figure 7.** The Mitrion-C data dependency graph for the source code shown in Code 3 from the Appendix.

Note that even though our Mitrion-C implementation of the computational kernel is setup as a cross-correlation subroutine, we also use this kernel to compute the autocorrelation function. The only drawback of using this subroutine to compute the autocorrelation is that the final bin counts need to be divided by two and the overall execution time is twice the time actually required to compute the autocorrelation. This approach was necessary since, at the time this work was done, Mitrion-C did not provide an efficient way to implement variable length loops as required for the inner loop in a true autocorrelation implementation.

Since the data structure used in the reference C implementation is not compatible with the required memory usage model shown in Tables 2 and 3, the data on the host system had to be reformatted before it was sent to the RC100 module memory. This adds to the total algorithm execution time, but the overhead was minimal. Code 4 in the Appendix shows the host wrapper pseudo code that demonstrates how the FPGA-accelerated code was used in the final implementation. As with the reference C implementation, since there are two FPGAs in the RC100 blade, the cross- and autocorrelation kernels were executed simultaneously, one subroutine per FPGA.

During the code compilation stage, the Mitrion-C compiler outputs various code analysis statistics, such as the bandwidth, execution time, read/write access to external memory, and the FPGA resources utilization. Thus, the overall Mitrion-C code execution time, as reported by the Mitrion-C compiler, was 94 seconds (9,440,259,920 steps at 100MHz), with 29% Flip Flops, 8% BlockRAMs, and 50% hardware multipliers used (see Table 4). These estimates are fairly close to the resource utilization statistics reported by the Xilinx place and route tools: 28% Flip Flops, 8% BlockRAMs, and 50% hardware multipliers usage. Overall, 47% of slices were occupied by the final design and all timing constraints were met. Execution time of a single run of the computational kernel, including data reformatting and transfer, was 99.4 seconds. Overall execution time of the entire program, including data I/O, was 10,071 seconds. Thus, we achieved an 8.6x overall code execution speedup as compared to the reference C implementation.

| | Total available | Mitrion-C estimate | After place and route |
|---|---|---|---|
| Slices | 89,088 | | 42,346 (47%) |
| Flip Flops | 178,176 | 51,327 (29%) | 50,128 (28%) |
| LUTs | 178,176 | | 48,724 (27%) |
| BlockRAMs | 336 | 29 (8%) | 27 (8%) |
| DSP48s | 96 | 48 (50%) | 48 (50%) |
| P&R time | 5 hours 24 minutes 7 seconds | | |

**Table 4.** The FPGA resource utilization as predicted by the Mitrion-C compiler (column 3) and the actual usage after the place and route (last column).

### 5.3. Kernel optimizations

In the previous section, a straightforward port of the computational kernel to RC100 platform was described. However, in order to realize the full potential of this platform one needs to consider how to better utilize all available resources and the added flexibility provided by the FPGA technology. In this section, we describe the various code optimization methodologies that were applied with the goal of achieving the highest possible performance of the algorithm.

#### 5.3.1. Multiple execution pipelines

In the first implementation, 47% of slices and 50% hardware multipliers were occupied by the final design. The Mitrion-C compiler uses 16 hardware multipliers to implement a single double-precision floating-point multiplier. This requires 48 (50% of all available) hardware multipliers to implement a 3D dot product on the Xilinx Virtex-4 VLX200 chip, leaving 48 hardware multipliers unused. As a result, there were sufficient FPGA resources left on the chip to implement an additional compute engine per chip.

The modifications to the Mitrion-C source code necessary to implement the two compute engines per chip were trivial: two points are loaded instead of one from the off-chip memory on each iteration of the outer loop and two separate dot product/bin mapping/bin update paths were instantiated inside the inner loop. The results were merged at the end before they are stored in the off-chip memory. No modifications to the previously used data storage or subroutine call from the host processor were required.

The overall Mitrion-C code execution time of the dual-kernel implementation, as reported by the Mitrion-C compiler, was 47.2 seconds (4,720,129,960 steps at 100MHz), with 45% Flip Flops, 8% BlockRAMs, and 100% hardware multipliers used (see Table 5).

| | Total available | Mitrion-C estimate | After place and route |
|---|---|---|---|
| Slices | 89,088 | | 70,301 (78%) |
| Flip Flops | 178,176 | 80,490 (45%) | 82,666 (46%) |
| LUTs | 178,176 | | 83,600 (46%) |
| BlockRAMs | 336 | 29 (8%) | 27 (8%) |
| DSP48s | 96 | 96 (100%) | 96 (100%) |
| P&R time | 1 days 11 hours 3 minutes 23 seconds | | |

**Table 5.** The FPGA resource utilization for dual-kernel design as predicted by the Mitrion-C compiler (column 3) and the actual usage after the place and route (last column).

The Xilinx place and route tools report 46% Flip Flops, 8% BlockRAMs, and 100% hardware multipliers usage, which was in a good agreement with the Mitrion-C compiler estimates. Overall, 78% of slices were occupied by the final design and all timing constraints were met. Execution time of a single run of the computational kernel, including data reformatting and transfer, is 49.7 seconds and the overall code execution time is 5,052.5 seconds – a 17x speedup as compared to the reference C implementation.

#### 5.3.2. Fixed-point, bit-width data

A closer examination of the numerical range of the bin boundaries revealed that only 12 digits after the decimal point (41 bits of the mantissa) are sufficient to cover our required numerical resolution. Thus, instead of using double-precision floating-point arithmetic, we could simply use fixed-point arithmetic with 42 bits allocated to store the absolute value and one bit allocated to store the sign. (Mitrion-C natively supports float and integer data types with an arbitrary number of bits, up to a limit.) This would result in significant savings in both the number of hardware multipliers and the logic required to implement the arithmetic operations. In the end, these resource savings could allow us to place four compute engines per chip instead of the initial two, thus potentially doubling the performance of the computational kernel. However, even though there were sufficient resources on the chip to place four computational kernels the design never met timing requirements, despite numerous attempts. A design with only three computational kernels using 42-bit integer arithmetic also did not meet timing requirements. A complete analysis of these results, however, is beyond the scope of this current paper.

#### 5.3.3. Optimization of the autocorrelation subroutine

As was mentioned in Section 5.3, the same FPGA-based computational kernel was used to compute both the cross-correlation and autocorrelation. In practice, this results in an inefficient implementation for the autocorrelation calculation as it requires twice the number of actual operations. When this work was performed, Mitrion-C did not provide an efficient way to implement variable length loops as required for the inner loop in the autocorrelation algorithm. However, since the combined number of outer/inner loop iterations is known: $N_D^2$ iterations in the case of the cross-correlation and $(N_D(N_D-1)/2)$ iterations in the case of the autocorrelation, the inner/outer loops can be fused into a single loop. As the result, a more efficient implementation of the autocorrelation kernel can be obtained. We did not pursue this implementation, however, as it would have little effect on the overall code performance, because in the present implementation the overall code performance is bound by the cross-correlation subroutine – the longer of two concurrent code sections.

### 5.3.4. Wide scaling and streaming

The RASC Abstraction Layer and the underlying RASC Core Services provide support for two unique capabilities: *multi-buffering* and *wide scaling*. In the case of wide scaling, the application can reserve all available devices and the Abstraction Layer can first spread the data across multiple devices and then loop over the devices until all the data is processed. In the case of multi-buffering, the Abstraction Layer divides data into smaller segments and overlaps the data transfer with the calculations on the FPGA. Multi-buffering can hide the data transfer cost, while wide scaling allows an efficient and fully transparent use of multiple FPGAs by the application.

Unfortunately, neither of these capabilities can be exploited in the current application. In order to enable the data streaming mode, each logical memory bank must be declared for either read or write access, but not for both. However, in our case one memory bank is declared for both read ($z$ coordinates of the points as shown in Tables 2 and 3) and write (bin counts are transferred out via this bank as well). This precludes the implementation of the double buffering mechanism required for the data streaming mode. Yet, even if it would have been possible to implement this capability, the savings would have been minimal as the FPGA algorithm is largely dominated by the computational kernel rather than transferring of data.

In the case of our system, there are two FPGAs available, but because both of them are used simultaneously for different tasks, the wide scaling capability, as provided by the RASC Abstraction Layer, cannot be used either. Note, however, that in our application the wide scaling capability is, in essence, setup and used explicitly: once the workload for each chip is defined, a separate thread is created manually.

## 5.4. On the fairness of the speedup comparison

Is it fair to compare the performance of our reference C implementation with the performance of the RC100 implementation? The reason why this question is brought up here is because the reference C implementation is setup to operate on an array of bins (into which the distances are mapped) of an arbitrary size whereas the Mitrion-C code is written to operate with the fixed size bin array. If we fix the size of the bin array in our reference C implementation and manually unroll the binary search loop, as is done by the compiler in the Mitrion-C implementation, the performance comparison will be more accurate since it will more accurately reflect the fact that both codes are optimally written to perform under similar conditions. Such an *optimized* implementation outperforms our original reference C implementation almost by a factor of 2 (see Table 6).

## 5.5. Performance results

Table 6 provides execution time measurements for the reference C implementation (column 2), two implementations in which the computational subroutine was successfully ported to the FPGA (columns 3 and 4), an estimation for the quad-kernel implementation (column 5) and the optimized reference C implementation (last column) as described in Section 5.4. FPGA execution time (row 4) is taken from the Mitrion-C compiler report while DD (autocorrelation for the actual dataset), DR+RR (autocorrelation for the random data plus cross-correlation for the actual and random data) and the load/convert times are actual measurements. Overall speedup is reported as the ratio between the total time spent by the **optimized reference C implementation** and the time spent by the **corresponding FPGA-accelerated implementation**. Note that the computational core was executed simultaneously by the two microprocessors in both the reference C implementation and the optimized reference C implementation and by the two FPGAs in the FPGA-accelerated implementations. Also, data I/O and data pre-processing stages are all included in the overall execution time. Thus, the end-to-end application speedup is reported rather than the computational core-only speedup, as often found in the literature. Our most successful implementation (Table 6, column 4) outperforms the optimized reference C implementation by a factor of 9.5. The integer arithmetic, quad-kernel implementation (Table 6, column 5) has a potential to outperform the microprocessor implementation by a factor of 18.8, however this design did not meet timing requirements.

| Measured features/ parameters | Reference C implementation | Code from section 5.2 | Code from section 5.3.1 | Code from section 5.3.2* | Optimized reference C |
|---|---|---|---|---|---|
| # CPUs | 2 | | | | 2 |
| # FPGAs | | 2 | 2 | 2 | |
| FPGA exec (s) | | 94 | 47.2 | 23.6 | |
| DD time (s) | 428.4 | 99.4 | 49.7 | *24.9* | 226.6 |
| DR+RR time (s) | 85,859.9 | 9,943.6 | 4,975.3 | *2,487.7* | 47,598.6 |
| Load/convert (s) | 27.4 | 28 | 27.5 | *28* | 28.4 |
| Total (s) | 86,315.7 | 10,071 | 5,052.5 | *2,540.5* | 47,853.6 |
| Overall Speedup | 0.6x | 4.8x | 9.5x | *~18.8x* | |

**Table 6.** The performance results for the three implementations. *) This column is an estimate for the quad-kernel implementation.

## 6. Discussion

In this section, we discuss specific issues that we uncovered while implementing the TPACF algorithm on the SGI RC100 compute blade.

An RC100 blade uses two Xilinx Virtex 4 VLX200 chips. These are large FPGAs; each chip contains 89,088 logic slices and 6,048 kbits of total Block RAM. However, VLX200 chips have only 96 18x18 hardware multipliers (called DSP48 slices), which is sufficient to implement only 6 double-precision floating-point multipliers using Mitrion-C. This is even less than what was available on the FPGAs used in the second-generation RASC boards [2]. Of course additional multipliers can be made out of logic cells, however, they require a substantial number of slices. Thus, while suitable for applications that require only a few floating-point multiplications, VLX200 chips are not appropriate for scientific computing applications that rely on double-precision floating-point operations.

Place and route times for large FPGAs, such as VLX200, are substantial. A design that occupies roughly 50% of the chip (see, e.g., Section 5.2) requires over 5 hours to place and route. Larger designs, for example, those that approach 100% slice utilization, result in lengthy bitstream synthesis. As an example, place and route time for the design presented in Section 5.3.1 is over 35 hours.

Each FPGA on the RC100 board is connected to five QDR SRAM DIMMs. However, the current version of the RASC Core Services provides access to only four of them for the user algorithm, arranged as two 128-bit wide banks. Thus, while the actual hardware provides some substantial bandwidth to the off-chip memory, the logical memory partitioning is not optimal for many applications, and it can actually prevent maximum performance from being achieved. Consider our application as an example. On each clock cycle, the computational core requires three words, each 64 bits wide. To be efficient, the data have to be spread between the two logical banks (see Section 5.2 for details). Thus, 2x128 bits of storage are required to store 3x64 bits of data. (Note that it is possible to compact the data so that no space is wasted, but it still will have to be stored in two logical memory banks to be efficient.) At the end, one of the logical memory banks is reused to send the results back to the host application. Thus, this memory bank is declared as read-write, which means that the double-buffering schema necessary to overlap the data movement with calculations cannot be implemented (we note that it would not affect the results for our application). However, if instead of having two 128-bit wide logical off-chip memory banks, the system would have provided four 64-bit wide memory banks, it would have been possible to use three such memory banks to send data in and one bank to send the results back to the host system, thus enabling us to implement the double-buffered streaming mode. An off-chip memory layout in which there are multiple 64-bit wide memory banks is, in general, more advantageous than the memory layout used in RC100.

Even though there are two FPGAs in the RC100 blade, no direct communication or data exchange path exists, at least as far as the user algorithm block in the FPGA is concerned. We would have preferred having the FPGAs in the blade being able to access the same off-chip common memory, thus enabling multi-stage data processing. If available though, this feature is unlikely to be of any use in our present test case application.

The RASC Abstraction Layer provides a unique wide scaling capability that enables multiple FPGAs configured with the same bitstream to work concurrently on the same problem by automatically dividing the workload between them. This functionality, however, can be trivially mirrored manually, as we did in our test application. Moreover, additional flexibility can be gained when using the manual procedure. The RASC library provides a straightforward way for allocating individual FPGAs for specific tasks and multiple FPGAs can be used simultaneously in separate threads. A RASC library wrapper code is simple and straightforward to write.

The RASC Abstraction Layer provides an ability to overlap data transfer and calculations via multi-buffering. This feature, however, comes with some restrictions on the type of data movement between the host memory and the FPGA memory, which means that not every application can take advantage of this capability. Problems that are "streaming" in nature and have no data dependencies can benefit from these modes of operation. However, iterative applications, such as our test case, do not gain much as they require a substantial amount of data reuse.

The GNU debugger (gdb) has been extended by SGI to enable debugging of the process executed in the FPGA(s). gdb interacts with the RASC Abstraction Layer to implement several debugging mechanisms, such as stepping through the execution on the FPGA and looking at the FPGA variables. We did not explore using the debugger in this research.

The Mitrion SDK for the RC100 has proven to be an effective tool to develop FPGA-accelerated algorithms. The compiler generates all the necessary configuration files, sets up the complex top-level project and compilation environment, and invokes the downstream compilation process for the RC100 RASC platform – a one button solution from the Mitrion-C source to the FPGA configuration bitstream. There is a considerable entry cost though as the semantics of the Mitrion-C language is rather different from C/C++ or FORTRAN – the languages of choice in high-performance computing. Once understood, however, the language is quite effective.

One major deficiency we ran into with the Mitrion SDK is the inability of the compiler to efficiently implement loops with a run-time defined number of iterations. Such loops can be implemented via the 'while' construct, however they are not pipelined by the compiler and therefore are not efficient. Future releases of the Mitrion SDK are expected to address his issue.

The RASC Abstraction Layer defines eight, 64-bit wide software-write/hardware-read registers that can be used to pass user arguments to the algorithm implemented on the FPGA. In its present implementation, the Mitrion SDK for RC100 does not provide support for these registers.

While using 42-bit multipliers, it was observed that the Mitrion-C compiler overestimated the need for FPGA hardware multipliers. For example, according to the Mitrion-C compiler, the three-kernel version of the code required 81 hardware multipliers, while, once placed and routed, the design required only 54 hardware multipliers.

As seen from Table 6, the FPGA execution time reported by the Mitrion-C compiler and the cross-correlation subroutine time measured during the actual execution differs by several seconds. For example, the Mitrion-C compiler reports 94 seconds as the algorithm execution time while we measure 99.4 seconds (Section 5.2) — a 5.4 second difference. As another example, the Mitrion-C compiler reports 47.2 seconds, while we measure 49.7 seconds (Section 5.3.1) — a 2.5 second difference. In both cases, the same host wrapper code is used (the same data preprocessing step and the same amount of data transferred in and out). The discrepancy between the compiler-reported and run-time-measured times for these two implementations differs by a factor of ~2, which is proportional to the overall algorithm execution time ratio. Such a discrepancy can be attributed to the compiler not being able to take into account the latency of the inner loop, or to the pipeline stalls that cannot be foreseen at the code compilation stage.

## 7. Conclusions

This evaluation, although limited to an application that requires double-precision floating-point arithmetic, shows that the SGI RC100 reconfigurable application-specific computing platform is a viable HPRC platform. Programming the RC100 using the Mitrion SDK is similar to programming other HPRC systems in the sense that the same code development methodology can be applied and the architecture and software-imposed challenges and limitations are similar to those found on other platforms. The Mitrion SDK for the RC100 has proven to be a workable solution to program the system, yet some language and compiler improvements, particularly with loops of variable numbers of iterations, are highly desirable. Some modifications of the RASC Core Services, particularly with the off-chip memory layout, would also result in a more robust platform.

## 8. Acknowledgements

This work was funded in part by the National Science Foundation grant SCI 05-25308 and by NASA grant NNG06GH15G. An SGI Altix 350 system with an RC100 blade was kindly made available to us by SGI. Mitrion SDK for RC100 RASC system was kindly provided to us by Mitrionics AB. We would like to thank Tony Galieti, Matthias Fouquet-Lapar, and Dick Riegner, all from SGI, for their help and support with the SGI system. Special thanks for Stefan Möhl and Jace Mogill, both from Mitrionics AB, for their help with Mitrion SDK. Special thanks for Dr. Craig Steffen from NCSA for useful comments.

## 9. References

[1] M.B. Gokhale, and P.S. Graham, *Reconfigurable Computing: Accelerating Computation with Field-Programmable Gate Arrays*, Springer, Dordrecht, 2005.

[2] Silicon Graphics Inc., Mountain View, CA, *Reconfigurable Application-Specific Computing User's Guide*, 2005.

[3] Silicon Graphics Inc., Mountain View, CA, *Reconfigurable Application-Specific Computing User's Guide*, 2006.

[4], Stefan Möhl, *The Mitrion-C Programming Language*, 2005, Mitrionics Inc., Lund, Sweden.

[5] A.D. Myers, R.J. Brunner, G.T. Richards, R.C. Nichol, D.P. Schneider, D.E. Vanden Berk, R. Scranton, A.G. Gray, and J. Brinkmann, "First Measurement of the Clustering Evolution of Photometrically Classified Quasars", The Astrophysical Journal, 2006, 638, 622.

[6] Silicon Graphics Inc., Mountain View, CA, *SGI Altix 350 System User's Guide*, 2004.

[7] S.D. Landy, and A.S. Szalay, "Bias and variance of angular correlation functions", The Astrophysical Journal, 1993, 412, 64.

[8] V. Kindratenko, A. Myers, R. Brunner, "Exploring coarse- and fine-grain parallelism on a high-performance reconfigurable computer", submitted to Parallel Computing.

## 10. Appendix

http://www.ncsa.uiuc.edu/~kindr/papers/FCCM07-appendix.pdf

```
for (i = 0; i < ((autoCorrelation) ? n1-1 : n1); i++)
{
   xi = data1[i].x;
   yi = data1[i].y;
   zi = data1[i].z;

   for (j = ((autoCorrelation) ? i+1 : 0); j < n2; j++)
   {
      double dot = xi * data2[j].x + yi * data2[j].y + zi * data2[j].z;

      // binary search
      min = 0;  max = nbins;
      while (max > min+1)
      {
         k = (min + max) / 2;
         if (dot >= binb[k])  max = k;
         else min = k;
      };

      if (dot >= binb[min])  data_bins[min] += 1;
      else if (dot < binb[max]) data_bins[max+1] += 1;
      else data_bins[max] += 1;
   }
}
```

**Code 1.** Pseudo code of the autocorrelation/cross-correlation computational kernel.

```
// compute bin boundaries
for (k = 0; k < nbins+1; k++)
   binb[k] = cos(pow(10,log10(min_arcmin) +
                   k*1.0/bins_per_dec) / 60.0*d2r);

// read data file
npd = readdatafile(file_name, data, npoints);

// compute DD
autoCorrelation(data, npd, DD, nbins, binb);

// loop through random data files
for (rf = 0; rf < args.random_count; rf++)
{
   // read random file
   npr = readdatafile(fname[rf], random, args.npoints);
   // compute RR and DR in parallel
   {
     autoCorrelation(random, npr, RRS, nbins, binb);
      crossCorrelation(data, npd, random, npr, DRS, nbins, binb);
   }
}

// compute TPACF
for (k = 1; k < nbins+1; k++)
   if (RRS[k] != 0) w[k] = (100.0 * DD[k] - DRS[k]) / RRS[k] + 1;
```

**Code 2.** Pseudo code of the overall algorithm.

```
Mitrion-C 1.0;
// options: -cpp
#define ExtRAM mem bits:128[262144]
#define NPOINTS 97160
#define NPOINTS_1 97159
```

```
#define NBINS 32

(ExtRAM, ExtRAM) main (ExtRAM a0, ExtRAM b0)
{
   float:53.11[NBINS] binb = [ 0.99999999999576916210,
         0.99999999998937272316, 0.99999999997330546453,
         0.99999999993294641509, 0.99999999983156895311,
         0.99999999957692020658, 0.99999999893727176126,
         0.99999999733054723006, 0.99999999329463784559,
         0.99999998315689186956, 0.99999995769202532081,
         0.99999989372717357217, 0.99999973305473655039,
         0.99999932946385972077, 0.99999831568965213968,
         0.99999576920548627346, 0.99998937273599586284,
         0.99997330559122843407, 0.99993294712784397404,
         0.99983157364604546835, 0.99957695008220059929,
         0.99893745993583771270, 0.99733173469789593302,
         0.99330212814615548300, 0.98320412044889693437,
         0.95798951231548890028, 0.89559620527121441835,
         0.74472199710330100331, 0.40112945079596057374,
        -0.26150795041456115220, -0.97304487057982380627,
        -100.0 ];

   uint:64[NBINS] binsA = [ 0, 0, 0, 0, 0, 0, 0, 0, 0, 0, 0, 0, 0, 0, 0,
                        0, 0, 0, 0, 0, 0, 0, 0, 0, 0, 0, 0, 0, 0, 0, 0, 0 ];

   ExtRAM a3 = a0;
   ExtRAM b3 = b0;

   // loop in one data set
   (bins, afinal, bfinal) = for (i in <0 .. NPOINTS_1>)
   {
      (xi, yi, zi, a1, b1) = readpoint(a0, b0, i); // read next point

      uint:64[NBINS] binsB = binsA;
      ExtRAM a2 = a0;
      ExtRAM b2 = b0;

      (binsA, a3, b3) = for(j in <0 .. NPOINTS_1>)
      {
         // read next point
         (xj, yj, zj, a2, b2) = readpoint(a1, b1, j+NPOINTS);

         // compute dot product
         float:53.11 dot = xi * xj + yi * yj + zi * zj;

         // find what bin it belongs to
         int:8 indx = findbin(dot, binb);

         // update bin
         binsB = foreach (bin in binsB by ind)
                  if (ind == indx) bin + 1 else bin;
      } (binsB, a2, b2);
   } (binsA, a3, b3);

   ExtRAM bw = b0;
   int:64 idx = 0;
   int:64<NBINS> binsR = reformat(bins, <NBINS>);

   bfinal2 = for (o in binsR)      // write results back to CPU
   {
      bits:128 out_val = [ idx, o ];
      bw = _memwrite(bfinal, idx, out_val);
      idx = idx + 1;
   } bw;
   bdone2 = _wait(bfinal2);
} (afinal, bdone2);

// given a value (dot) and a vector of bin boundaries (binb),
// return bin index the given value belongs to
```

```
(int:8) findbin(float:53.11 dot, float:53.11[NBINS] binb)
{
    int:8 asas = 0;
    bool found = false;
    int:8 res = 0;

    int:8 indx = for (b in binb)
    {
        (res, found) = if (dot >= b && !found)
        {
            found = true;
        } (asas, found)
        else {} (res, found);

        asas = asas + 1;
    } res;
} indx;

// read one point value from the external memory
(float:53.11, float:53.11, float:53.11, ExtRAM, ExtRAM)
readpoint(ExtRAM a0, ExtRAM b0, int:64 indx)
{
    (val1, a1) = _memread(a0, indx);  // read x,y
    (val2, b1) = _memread(b0, indx);  // read z

    float:53.11[2] via1 = val1;      // unpack
    float:53.11[2] via2 = val2;

    float:53.11 x = via1[0];
    float:53.11 y = via1[1];
    float:53.11 z = via2[0];
} (x, y, z, a1, b1);
```

**Code 3.** Mitrion-C implementation of the cross-correlation computational kernel.

```
// allocate memory for BRAM
char a0_in[bram_size], b0_in[bram_size], b0_out[bram_size];
long long *bin_val = (long long*)b0_out;

// copy first dataset
for (i = 0; i < n1; i++) {
    double *xi = (double*)(a0_in+i*16);
    double *yi = (double*)(a0_in+8+i*16);
    double *zi = (double*)(b0_in+i*16);
    *xi = data1[i].x;
    *yi = data1[i].y;
    *zi = data1[i].z;
}
// copy second dataset
for (j = 0; j < n2; j++) {
    double *xj = (double*)(a0_in+j*16 + sizeof(double)*n1*2);
    double *yj = (double*)(a0_in+8+j*16 + sizeof(double)*n1*2);
    double *zj = (double*)(b0_in+j*16 + sizeof(double)*n1*2);
    *xj = data2[j].x;
    *yj = data2[j].y;
    *zj = data2[j].z;
}
// do FPGA work
rasclib_cop_send(algorithm_id, "a0_in", a0_in, bram_size);
rasclib_cop_send(algorithm_id, "b0_in", b0_in, bram_size);
rasclib_cop_go(algorithm_id);
rasclib_cop_receive(algorithm_id, "b0_out", b0_out, bram_size);
rasclib_cop_commit(algorithm_id, NULL);
rasclib_cop_wait(algorithm_id);

// copy data out, and apply fix for autocorrelation
for (k = 0; k < nbins+2; k++)
    data_bins[k] += (autoCorrelation) ? ((k==0) ? 0 :
                          bin_val[2*k+1]/2) : bin_val[2*k+1];
```

**Code 4.** Host wrapper pseudo code.